\documentclass[prl,aps,twocolumn]{revtex4}
\usepackage{epsfig}
\usepackage{bm}

\usepackage[latin1]{inputenc}
\newcommand{\beq}{\begin{equation}}
\newcommand{\eeq}{\end{equation}}
\newcommand{\bea}{\begin{eqnarray}}
\newcommand{\eea}{\end{eqnarray}}
\begin{document}
\title{Fracture Surfaces as Multiscaling Graphs}
\author{Eran Bouchbinder$^1$, Itamar Procaccia$^{1}$, St\'ephane Santucci$^2$ and Lo\"ic Vanel$^2$}
\affiliation{$^1$Dept. of Chemical Physics, The Weizmann Institute
of Science, Rehovot 76100, Israel\\
 $^2$Laboratoire de physique, CNRS UMR 5672, Ecole Normale Sup\'erieure de Lyon,
46 all\'ee d'Italie, 69364 LYON Cedex 07, France}

\begin{abstract}
Fracture paths in quasi-two-dimensional (2D) media (e.g thin  layers
of materials, paper) are analyzed as self-affine graphs $h(x)$ of
height $h$ as a function of length $x$. We show that these are
multiscaling, in the sense that $n^{th}$ order moments of the height
fluctuations across any distance $\ell$ scale with a characteristic
exponent that depends nonlinearly on the order of the moment. Having
demonstrated this, one rules out a widely held conjecture that
fracture in 2D belongs to the universality class of directed
polymers in random media. In fact, 2D fracture does not belong to
any of the known kinetic roughening models.  The presence of
multiscaling offers a stringent test for any theoretical model; we
show that a recently introduced model of quasi-static fracture
passes this test.
\end{abstract}
\maketitle

{\bf Background}: The pioneering experimental work described in Ref.~\cite{84MPP} drew attention to the fact that fracture
surfaces are  graphs in 2+1 (1+1) dimensions when the broken sample is three dimensional (two dimensional).
This initial insight was followed up by a considerable number of works \cite{03Bou} that focused on
the scaling properties of such
graphs under affine transformations. In 1+1 dimensions one denotes the graph as
$h(x)$ and considers the structure function $S_2(\ell)$,
\begin{equation}
 S_2(\ell)\equiv \langle |h(x+\ell)-h(x)|^2\rangle \ , \label{S2}
 \end{equation}
 where angular brackets denote an average over all $x$. The statement is then that this function
 is a homogeneous function of its arguments,
 \begin{equation}
 S_2(\lambda \ell)\sim \lambda^{\zeta_2} S_2(\ell) \ . \label{zeta2}
 \end{equation}
 Close attention was paid to the numerical value of the scaling exponent $\zeta_2/2$ as observed in rupture lines in
 quasi two-dimensional materials (and in some numerical simulations) \cite{93KHW, 94EMHR, 03SAN, 91HHR}
 with the hope of assigning to fracture
 a ``universality class" of one of the well studied models of 1+1 dimensional kinetic roughening
 models \cite{95H-HZ,95BS}. Indeed, since in many such 1+1 measurements the exponent $\zeta_2/2$ was numerically close to 0.67,
 and also since it was proposed that fracture can be considered as a global minimization problem,
 some authors accepted a view that rupture lines map onto the model of directed polymers in random media
 \cite{91HHR, 93KHW, 95H-HZ, 95BS}.
 The latter model is consistent with an exactly soluble exponent $\zeta_2/2=2/3$ \cite{95BS}.
 In this Letter we show that this is not the case;
 moreover we advance reasons to believe that rupture lines do not map on any of the known 1+1 kinetic roughening models,
 but call for fresh thinking with new models in mind.
 One such new model will be shown to be a good candidate for this type of physics.

{\bf Approach}: We first argue in this Letter that the scaling properties of fracture lines are significantly
richer than what can be read from Eq. (\ref{zeta2}).
To this aim we consider the whole distribution function for height fluctuations, $P\Big(h(x+\ell)-h(x)\Big)$,
and the associated higher order structure function $S_n(\ell)$,
\begin{equation}
 S_n(\ell)\equiv \langle |h(x+\ell)-h(x)|^n\rangle \ . \label{Sn}
 \end{equation}
For rupture lines the higher order structure function come each with its own exponent,
\begin{equation}
S_n(\lambda \ell)\sim \lambda^{\zeta_n} S_n(\ell) \ , \label{scaling}
\end{equation}
 where the scaling exponents $\zeta_n$ are not simply related to the exponent $\zeta_2$ (i.e.
$ \zeta_n\ne n\zeta_2/2$).
In contrast, in directed polymers in random media the higher order structure functions bring in no new information,
and there $ \zeta_n= n\zeta_2/2$ \cite{91Hal}.

{\bf Experimental example}: An example that provides us with
information of sufficient accuracy to establish the multiscaling
characteristics is rupture lines in paper. The data acquired by
Santucci et al. \cite{04SVC} was obtained in experiments where
centrally notched sheets of fax paper were fractured by standard
tensile testing machine. Four resulting crack profiles $h(x)$ were
digitized. Each digitization contained a few thousand points, where
care was taken to insure that the smallest separation between points
in $x$ is larger than the typical fiber width; this is important to
avoid the artificial introduction of overhangs that destroy the
graph property.

\begin{figure}[here]
\centering
\epsfig{width=.45\textwidth,file=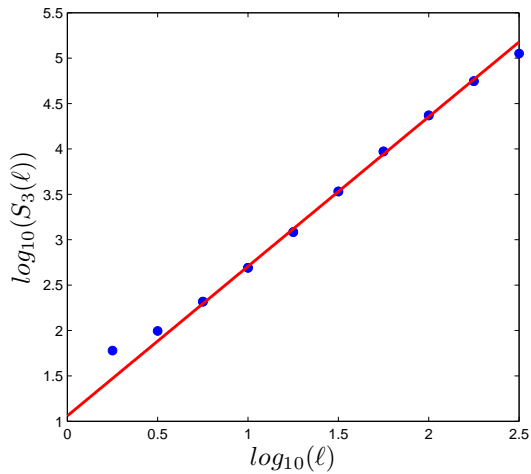}
\caption{A log-log plot of $S_3(\ell)$ as a function of $\ell$. The linear fit corresponds
to a typical scaling range of about $1.5$ orders of magnitude, with a slope of $\zeta_3 \simeq 1.65$.}
\label{Example}
\end{figure}

\begin{figure}[here]
\centering
\epsfig{width=.45\textwidth,file=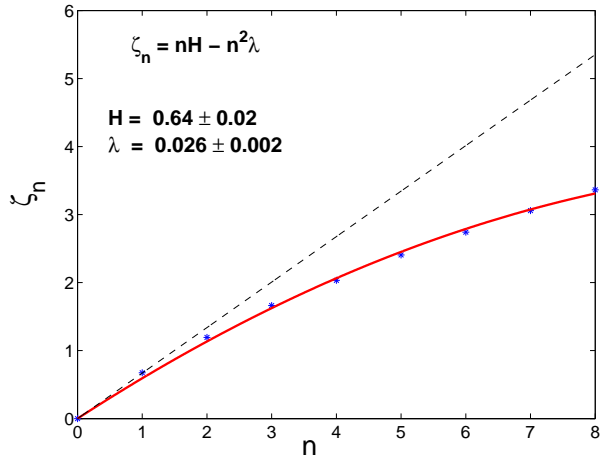}
\caption{The spectrum $\zeta_n$ as a function of the moment order
$n$ for rupture lines in paper. The function is fitted to the form
$\zeta_n = nH-n^2\lambda$ and the parameters $H$ and $\lambda$ are
given. The errors in the estimation of these parameters reflect both
the variance between different samples and the fit quality.
The linear plot $n\zeta_1$ is added to stress the non-linear nature of
$\zeta_n$.}\label{Multiaffine_spectrum}
\end{figure}

Denoting $\Delta h (\ell) \equiv h(x+\ell)-h(x)$ we analyzed the
data by boxing $\ell$ in logarithmic boxes, accumulating the data
between $10^0$ and $10^{0.25}$ (the smallest box) and between
$10^{2.25}$ and $10^{2.5}$ (the largest box). The $m^{th}$ box was
considered as representing data for $\ell=10^{m\times 0.25}$. On the
basis of this boxing we constructed the  probability distribution
function (pdf) $P(\Delta h(\ell))$ by combining data from all the
four samples. Samples that exhibit marked trends (probably due to
the finite size of the sample), were detrended by subtracting the
mean from each distribution. The computed pdf's were then used to
compute the moments (\ref{Sn}), and these in turn, once presented as
log-log plots, yield the scaling exponents $\zeta_n$. Such a typical
log-log plot is shown in Fig. \ref{Example}, exhibiting a
typical scaling range of about $1.5$ orders of magnitude. The
resulting values of the scaling exponents $\zeta_n$ are shown in
Fig. \ref{Multiaffine_spectrum}.

\begin{figure}[here]
\centering \epsfig{width=.475\textwidth,file=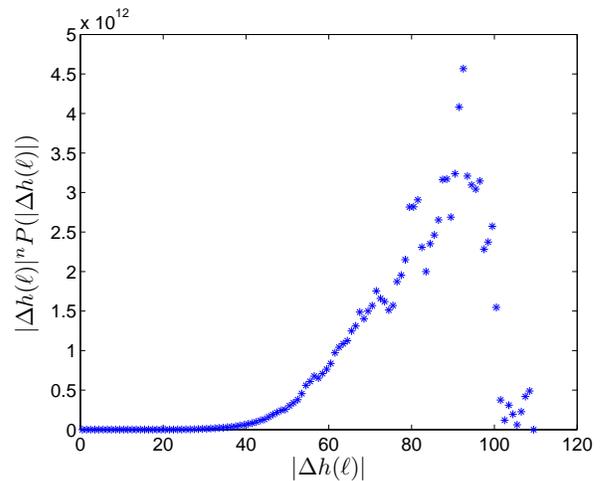}
\caption{An example of the convergence of the integral in Eq.
(\ref{integral}) for $\ell=10^{2.25}$ and $n=8$.} \label{Convergence}
\end{figure}

As a function of $n$ these numbers  can be fitted to the quadratic function
$\zeta_n = nH-n^2\lambda$ with $H = 0.64 \pm 0.02$ and $\lambda =
0.026 \pm 0.002$ (a linear plot $n \zeta_1$ is added for reference). The error bars quoted here reflect both the
variance between different samples and the fit quality.  The
exponents were computed for $n\le 8$ since for higher moments the
discrete version of the integral
\begin{equation}
\int |\Delta h(\ell) |^n P(\Delta h(\ell))d \Delta h(\ell) \label{integral}
\end{equation}
did not converge. On the other hand, the convergence of the $8^{th}$
order moment is demonstrated in Fig.~\ref{Convergence}.

\begin{figure}
\centering \epsfig{width=.48\textwidth,file=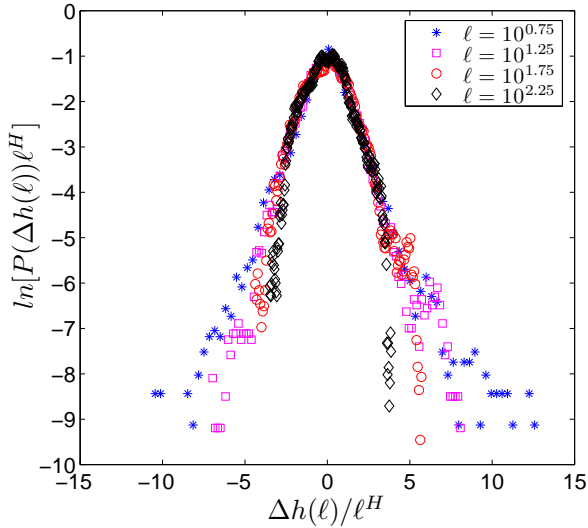}
\caption{The natural logarithm of $P(\Delta h(\ell))\ell^{H}$ as a
function of $\Delta h(\ell)/\ell^{H}$ for $H=0.64$. The legend gives the
scale $\ell$ for each plot.}\label{Data_Collapse}
\end{figure}

The point to stress is that the scaling exponents
$\zeta_n$ depend non-linearly on $n$; for the range of $n$ values for which
the moments converge, the exponents can be fitted to a
quadratic function. It is well known
from other areas of nonlinear physics, and turbulence in particular \cite{84AGHA,95Fri,05BP}, that
such phenomena of multiscaling are associated with pdf's on different
scales $\ell$ that cannot be collapsed by simple rescaling. In other words,
in the absence of multiscaling, there exists a single scaling exponent $H$
with the help of which one can rescale the pdf's according to
\begin{equation}
P(\Delta h(\ell)) \sim \ell^{-H} f\left(\frac{\Delta h(\ell)}{\ell^{H}}\right)\ ,
\label{distribution}
\end{equation}
where $f(\cdot)$ is a scaling function. In our case such rescaling does not
result in data collapse. In Fig.
\ref{Data_Collapse} the natural logarithm of $P(\Delta h(\ell))\ell^{H}$
is plotted as a function of $\Delta h(\ell)/\ell^{H}$ for $H=0.64$.
Indeed, the data does not collapse onto a single curve.
The fat tails of the probability distribution functions at
smaller scales is typical to multiscaling situations.

The known kinetic roughening models, and in particular directed polymers
in random media, do not exhibit the multiscaling spectrum we have found, and thus we can infer that
the scaling properties of rupture lines in 1+1 dimension do not fall in the
universality class of any one of the former. We propose that the long range
elastic interaction that is typical to fracture is the origin of the positive
correlations resulting in a non-trivial exponent (note that the quasi static fracture problem
involves the solution of the bi-Laplacian equation). To appreciate this important difference
from standard kinetic roughening models in 1+1 dimensions we turn now
to a similar analysis of a theoretical model.

{\bf Theoretical example}:  A recently published work \cite{04BMP, 05ABKMP} described a crack growth model
aiming at understanding quasi-static fracture via voids formation and coalescence.
The model consisted of
two components: (i) the exact solution of the elasticity problem in
the presence of an arbitrarily shaped crack (ii) a growth law in
which the evolution of the crack is controlled by the nucleation of
voids at a {\em finite distance} $R$ ahead of the crack tip. The
first component involved the adaptation of the method of iterated
conformal maps to the field of elasticity in complex geometrical
domains \cite{04BMPa} . The second component incorporates an important physical
insight about the role of a typical {\em length scale} $R$ in the
problem. This length scale determines where voids nucleate,  making
the crack growth a succession of rapid growth steps interrupted by slower
void nucleation events.  This is a crucial factor that
enables the crack to develop positive correlations (i.e. a roughness
exponent such that $H>1/2$) \cite{88Fed}. The succession of growth steps
introduce variations in the crack geometry on the scale of the step size $R$.
At a distance $R$ ahead of the crack tip these geometric irregularities  change the solution of the stress
field and  mediate positive correlations. When the growth
takes place right {\bf at} the crack tip \cite{97REF, 02BHLP}, the crack
appears locally straight;  the geometric irregularities
are effectively screened and no positive correlations appear, i.e.
one obtains $H=1/2$. The origin of the finite length scale that is
involved in the crack growth process can be the near tip non-linear
physics (for example plasticity, see \cite{04BMP, 05ABKMP}) or the scale
of the quenched disorder in the system. The existence of a finite
scale before the tip appears crucial for the development of
a non-trivial roughness exponent.

In \cite{04BMP, 05ABKMP} it was shown
that this particular growth model generates rupture lines with $\zeta_2/2
\approx 0.66$. It is thus interesting to examine whether this model exhibits the
same kind of multiscaling that was found in the experimental example above. Unfortunately,
due to the significant  computational cost of the iterated conformal maps technique
the numerical investigation of the growth model had a limited number of
realizations of a few hundreds growth steps. Due to the relative paucity of
data the structure functions defined in Eq. (\ref{Sn}) would not converge well enough
to provide reliable exponents. Instead, we use the max-min method which was proved reliable
for the range of exponents under study and for the typical length of our cracks \cite{95SVR}.
Thus, we define
\begin{equation}
\tilde{S}_n(\ell)\equiv \langle |Max\left\{h(\tilde x)\right\}_{x<\tilde x<x+\ell}-
Min\left\{h(\tilde x)\right\}_{x<\tilde x<x+\ell}|^n\rangle \ . \label{newS}
\end{equation}
The scaling exponents $\zeta_n$ are defined in analogous way to Eq. (\ref{scaling}) by
\begin{equation}
\tilde{S}_n(\lambda \ell)\sim \lambda^{\zeta_n} \tilde{S}_n(\ell) \ .
\end{equation}

\begin{figure}[here]
\centering \epsfig{width=.45\textwidth,file=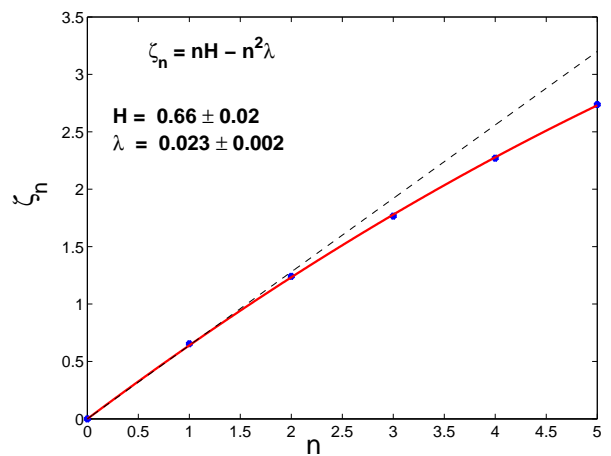}
\caption{The spectrum $\zeta_n$ as a function of the moment order
$n$ for rupture lines in the model of \cite{04BMP, 05ABKMP}. The function is fitted to the
form $\zeta_n = nH-n^2\lambda$ and the parameters $H$ and
$\lambda$ are given. The errors in the estimation of these
parameters reflect both the variance between different
realizations and fit quality. The linear plot $n\zeta_1$ is
added for comparison.}\label{model}
\end{figure}

The resulting exponents $\zeta_n$ for the cracks
generated by this model are shown in Fig. \ref{model}.
For the low orders moments (here we are limited by the paucity of data to
$n\le 5$) one again fits  a quadratic function, with $H =
0.66 \pm 0.02$ and $\lambda = 0.023 \pm 0.002$. The errors in
the estimation of these parameters reflect both the variance
between different realizations and the fit quality.
The $n$ dependence of the
exponents $\zeta_n$ and the values of the fitting parameters are
in agreement with the experimental ones. Since there is nothing in
the model that is specific for the physics of paper, it appears
that multiscaling is a generic property of the fracture process,
at least in 1+1 dimensions.

{\bf Summary}: We examined an experimental example of rupture lines and a theoretical
model  of rupture in 1+1 dimensions, and showed that both exhibit graphs whose scaling
properties appear similar, falling in a different class compared to standard models
like directed polymers
in random  media.  The scaling exponents of higher order
structure functions associated with graphs of rupture lines depend non-linearly on the
order of the moments, while in directed polymers  this is not the case. In fact, it appears that
none of the standard known growth models in 1+1 dimensions is in the same class as
the rupture lines discussed here. The analysis of the theoretical model appears encouraging,
in the sense that it captures the multiscaling characteristic of the experimental example. We reiterate
our proposition that the crucial aspect of the theoretical model that is responsible for the non-trivial
scaling behavior is the existence of a finite scale $R$ that does not exist in pure elasticity theory,
or in models that treat the crack dynamics as a continuous process.

Further experimental work is necessary to provide further examples of rupture lines in 1+1
dimensions. With more examples one would be able to determine the degree of universality
of the scaling exponents, the dependence of the multiscaling exponents on issues like
the isotropy of the cracking medium, the presence of plasticity and the protocol of the rupture
experiments.

{\bf Acknowledgments:} We thank S. Ciliberto for providing the extensive
experimental results on paper rupture and E. Katzav for digitizing the data.
This work had been supported in part by the Israel Science Foundation administered by the Israel Academy of Sciences.
 E.B. is supported by the Horowitz Complexity Science Foundation.

\end{document}